\documentclass[aps,prx,twocolumn,amsmath,amssymb,superscriptaddress,longbibliography]{revtex4-2}
\usepackage[english]{babel}
\usepackage{xcolor}
\usepackage{amssymb}
\usepackage{dcolumn}
\usepackage{bm}
\usepackage{graphicx}
\usepackage{amsmath}
\usepackage{braket}

\allowdisplaybreaks[4]   %Formula span

\usepackage{graphicx}        % standard LaTeX graphics tool

\graphicspath{{pict/}{}}

\usepackage[normalem]{ulem}
\usepackage{dcolumn}%new
\usepackage{bm}
\usepackage[pdfstartview=FitH, CJKbookmarks=true, bookmarksnumbered=true, bookmarksopen=true, colorlinks=true, pdfborder=001, citecolor=blue, linkcolor=blue, urlcolor=blue, linktocpage=true] {hyperref}

\begin{document}

\title{Hong-Ou-Mandel Interference in a temporal-average-inversion-symmetric chain}

\author{Shi Hu}
\email{hush27@gpnu.edu.cn}
\affiliation{School of Optoelectronic Engineering, Guangdong Polytechnic Normal University, Guangzhou 510665, China}

\author{Meiqing Hu}
\affiliation{School of Optoelectronic Engineering, Guangdong Polytechnic Normal University, Guangzhou 510665, China}

\author{Shihao Li}
\affiliation{School of Optoelectronic Engineering, Guangdong Polytechnic Normal University, Guangzhou 510665, China}

\author{Zihui Zhong}
\affiliation{School of Optoelectronic Engineering, Guangdong Polytechnic Normal University, Guangzhou 510665, China}

\author{Zhoutao Lei}
\email{leizht3@mail2.sysu.edu.cn}
\affiliation{Guangdong Provincial Key Laboratory of Quantum Metrology and Sensing $\&$ School of Physics and Astronomy, Sun Yat-Sen University (Zhuhai Campus), Zhuhai 519082, China}
\date{\today}

\begin{abstract}
We show how to implement tunable beam splitter and Hong-Ou-Mandel interference in the Su-Schrieffer-Heeger chain by manipulating the topological edge states adiabatically. The boson initially injected in the one end of the chain can be transferred to the two-end with a tunable proportion depends on the dynamical phases accumulated during the adiabatic evolution. We also observe Hong-Ou-Mandel interference via the tunable beam splitter ($50:50$) and achieve a spatially entangled two-particle NOON state. We demonstrate the robustness of our proposal under chiral- and time-reversal-symmetry-preserving disorder. However, the chiral symmetry is scarce for realist system. Therefore, we demonstrate Hong-Ou-Mandel interference are robust to inversion symmetric disorder breaking the chiral symmetry, highlighting the protection of inversion symmetry. More importantly, the inversion symmetry violated by static disorder can be restored for more common situations where disorder becomes time dependent, giving rise to the temporal-average-inversion-symmetry protected Hong-Ou-Mandel interference. Our approach opens a new way to study quantum effects in topological matter with potential applications.
\end{abstract}
\maketitle

%%%%%%%%%%%%%%%%%%%%%%%%%%%%%%%%%%%%%%%%%%%%%%%%%%%%%%%%%%%%%%%%%%%%%%%%%%%%%%%%%%%%
\section{Introduction}\label{Sec1}
%%%%%%%%%%%%%%%%%%%%%%%%%%%%%%%%%%%%%%%%%%%%%%%%%%%%%%%%%%%%%%%%%%%%%%%%%%%%%%%%%%%%
Topological states of matter, originally discovered in condensed matter~\cite{DXiaoRMP2010,MZHasanRMP2010,XLQiRMP2011}, have attracted intensive attention in many fields such as photonics~\cite{HaldanePRL2008,ZWangNature2009,LuNP2014,TOzawaRMP2019}, acoustics~\cite{ZYangPRL2015,MaNRP2019,XueNRM2022}, and cold atoms~\cite{MAidelsburgerPRL2013,CooperRMP2019} over the last decade. The existence of topologically protected edge states according to bulk-edge correspondence~\cite{YHatsugaiPRL1993,EssinPRB2011} is one of the most appealing characteristics of topological systems. The unique properties of the in-gap topological edge states such as robustness rooted in the bulk topological invariants~\cite{Ryu_2010,CKChiuRMP2016} and unidirectional propagation immune to backscattering~\cite{ZWangNature2009,JSeoNature2010} drive numerous research in quantum science and technology. These prominent features make topological edge states as reliable platform for topological protection of quantum correlation~\cite{ABlancoScience2018}, quantum state transfer~\cite{YEKrausPRL2012,NLangQI2017,CDlaskaQST2017,FMeiPRA2018,SLonghiPRB2019,NEPalaiodimopoulosPRA2021,LHuangPRA2022,CWangPRA2022}, topological quantum gates~\cite{PBorossPRB2019,MNarozniakPRB2021}, and topological quantum devices~\cite{RHammerPRB2013,XSWangPRB2017,LQiPRB2021,LQiQuantum2021,LQiPRA2023}. Actually, the characterization and properties of topological phases rely on their underlying symmetry. After establishing the celebrated Altland-Zirnbauer symmetry classification~\cite{PhysRevB.55.1142}, researchers began investing the role of spatial symmetry and introduce the concept of topological crystalline phases~\cite{PhysRevLett.98.106803,Slager2013,PhysRevX.7.041069,Bradlyn2017,Po2017,Elcoro2021}.

Quantum entanglement and interference play a crucial role in quantum communication, quantum computation, and quantum metrology~\cite{AEkertRMP1996,JWPanRMP2012,LPezzeRMP2018}. Topological systems have demonstrated topological protection of quantum entanglement~\cite{MCRechtsmanOptica2016,MWangNanophotonics2019,KMonkmanPRR2020,JXHanPRA2021}. A recent experiment~\cite{JLTambascoSA2018} has observed the well-known Hong-Ou-Mandel (HOM) interference~\cite{CKHongPRL1987} in an integrated photonic circuit described by off-diagonal Harper model~\cite{PGHarper1955}. In that work, topological interface was created in the middle of the system and interference was observed at this interface by adiabatically manipulating coupling strength and making topological edge modes into delocalized. Besides, one of us has demonstrated the HOM interference by Thouless pumping via topological bulk bands and generated spatially entangled two-particle NOON state ~\cite{SHuPRA2020}. It is interesting to study quantum interference and entanglement generation in topological system. 

In this work, we propose a scheme to achieve HOM interference and entanglement generation based on the Su-Schrieffer-Heeger (SSH) chain~\cite{WPSuPRL1979}, one of the most prototypical topological models. In topological nontrivial phase, the in-gap eigenstates of finite-sized SSH chain are odd and even superpositions of edge states localized exponentially on left and right edge~\cite{JKAsboth2016}. Moreover, edge states just localize in ends of chain in the fully dimerized limit (only intercell hopping). Firstly, we show a boson injected in one end of the chain, acting as a superposition of these two in-gap eigenstates, can be transferred to two ends with tunable proportion. This proportion depends on the dynamical phase accumulated in the adiabatic evolution, that is moving away from the fully dimerized limit and then return adiabatically. Therefore we have achieve a tunable beam splitter (BS) by manipulating the dynamical phase delicately, where two ends of SSH chain act as input and output ports. Extent it into two-particle system, HOM interference and generation of spatial two-particle NOON state can be observed via the tunable BS ($50:50$).

After studying the clean system with translation symmetry, we reveal the role of internal and inversion symmetries by introducing different type of disorder. More specifically, the chiral- and time-reversal-symmetry-preserving disorder only make a tiny random shift on the energy spectrum and influence the dynamical phase. In the static case with fixed disorder during the adiabatic evolution, we can adjust the total evolution time to match the desired dynamical phase. In the temporal case with time dependent disorder, the influence on the dynamical phase can be averaged out during the adiabatic evolution. However, the chiral symmetry is delicate for many realistic systems, such as the ones with on-site energy disorders. 
Therefore, we examine the effects of spatial symmetry and reveal the tunable BS and HOM interference still can be established against inversion-symmetric disorders, while can be destroyed by static generic ones. Fortunately, temporal-average-inversion-symmetry emerges for these cases with time dependent disorder, which exists universally in realistic systems, and then the BS as well as HOM interference is restored. These results highlight our proposal can be protected by temporal-average-inversion-symmetry, dual to previous works~\cite{PhysRevB.86.115112,PhysRevB.87.035119,LFuPRL2012,PhysRevB.89.155424,PhysRevResearch.2.012067,PhysRevLett.126.206404,PhysRevX.13.031016,Tao2023}  demonstrating the average spatial symmetry protected topological phase in statistic ensemble.

This paper is organized as follows. In Sec.~\ref{Sec2}, we introduce our model, SSH chain, its hybridized edge states, and underlying symmetries. In Sec.~\ref{Sec3}, we study how to achieve tunable BS ,HOM interference, and entanglement generation via edge channels of clean or disordered SSH chain remaining in class BDI. In Sec.~\ref{Sec4}, we investigate disordered systems without chiral symmetry and analyze the role of inversion symmetry, especially the case with temporal diagonal disorder where exact inversion symmetry are also broken while temporal-average-inversion-symmetry appears. In Sec.~\ref{Sec5}, we give a summary and discussion of our results.

%%%%%%%%%%%%%%%%%%%%%%%%%%%%%%%%%%%%%%%%%%%%%%%%%%%%%%%%%%%%%%%%%%%%%%%%%%%%%%%%%%%%
\section{SU-SCHRIEFFER-HEEGER MODEL AND HYBRIDIZED EDGE STATES}\label{Sec2}
\begin{figure}[!htp]
\includegraphics[width=0.6\columnwidth]{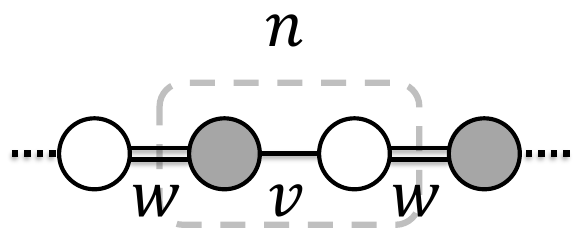}
\caption{\label{SSHModel} Schematic diagram of the SSH model. Filled (empty) circles are odd (even) sites, which are grouped into unit cells. The $n$-th cell is circled by a dotted line. $v$ and $w$ denote the intracell and intercell hopping, respectively.}
\end{figure}

%%%%%%%%%%%%%%%%%%%%%%%%%%%%%%%%%%%%%%%%%%%%%%%%%%%%%%%%%%%%%%%%%%%%%%%%%%%%%%%%%%%%
The SSH model is one of the most prototypic models exhibiting topological edge states. It describes particles hopping on a one-dimensional chain with staggered hopping amplitudes ($\hbar=1$):
\begin{eqnarray}\label{Ham} 
\hat{H}=v\sum_{n=1}^{N}\hat{b}_{2n-1}^{\dag}\hat{b}_{2n}
+w\sum_{n=1}^{N-1}\hat{b}_{2n}^{\dag}\hat{b}_{2n+1}+{\rm H.c.},
\end{eqnarray}
with $N$ unit cells (the total lattice site is $2N$). $\hat{b}_n^{\dag}$ and $\hat{b}_n$ are bosonic creation and annihilation operators acting on lattice site $n$, respectively. $v$ and $w$ denote the intracell and intercell hopping amplitudes, as shown in Fig.~\ref{SSHModel}. Under the periodic boundary condition, the Hamiltonian can be expressed in momentum space:
\begin{eqnarray}\label{Hamk}
\hat{H}=\sum_{k}\Psi_{k}^{\dag}H(k)\Psi_{k},~~
\Psi_{k}^{\dag}
=(\hat{b}_{{\rm odd},k}^{\dag}~~\hat{b}_{{\rm even},k}^{\dag}),\cr\cr
H(k)=(v+w\cos{k})\sigma_{x}+w\sin{k}\sigma_{y},~~~~~~~~
\end{eqnarray}
with $\sigma_{x,y,z}$ being Pauli matrices. For simplicity, we take $v$ and $w$ to be real and nonnegative. In following discussion, the intercell hopping is set as energy unit $w=1$.

SSH model holds time-reversal symmetry $H(k)=H^*(-k)$ (all hopping being real), particle-hole symmetry $\sigma_{z}H(k)\sigma_{z}=-H^*(-k)$ and chiral symmetry $\sigma_{z}H(k)\sigma_{z}=-H(k)$. Because both the time-reversal symmetry operator and particle-hole symmetry operator square to $+1$, SSH model falls into the BDI class in the Altland-Zirnbauer symmetry classification~\cite{PhysRevB.55.1142}. Moreover, SSH model also holds inversion symmetry read as $\sigma_{x}H(k)\sigma_{x}=H(-k)$. 
Either chiral symmetry~\cite{Ryu_2010,CKChiuRMP2016} or inversion symmetry~\cite{PhysRevB.89.155114,PhysRevA.102.013301,PhysRevB.103.024205} can protect nontrivial topological phases for SSH model when $v<w$. In this regime, single-particle SSH chain hosts two zero-energy edge states localized on the boundaries in the thermodynamic limit of $N\rightarrow\infty$, according to the well-known bulk-boundary correspondence. 

We now turn to consider a finite-sized system. The energies of the edge states remain very close to zero energy and take a pair of almost-zero-energy eigenvalues opposite to each other due to chiral symmetry. Actually, for an eigenstate state $|\psi_{n}\rangle$ with energy $E_n\neq0$, there is a chiral symmetric partner $\hat{S}|\psi_{n}\rangle$. Here $\hat{S}$ is the (single-particle) real space form of chiral symmetry operator expressed with the orthogonal projectors $\hat{P}_{{\rm odd}}$ and $\hat{P}_{{\rm even}}$:
\begin{eqnarray}\label{ChiralOperator} 
\hat{P}_{{\rm odd}}=\sum_{n=1}^{N}|2n-1\rangle\langle2n-1|,~~
\hat{P}_{{\rm even}}=\sum_{n=1}^{N}|2n\rangle\langle2n|,\cr\cr
\hat{S}=\hat{P}_{{\rm odd}}-\hat{P}_{{\rm even}},~~~~~~~~~~~~~~~~~~
\end{eqnarray}
and the eigenenergies of $\hat{S}|\psi_{n}\rangle$ will be $-E_n$ due to the chiral symmetric relationship (See Appendix~\ref{appA} for more details).

These almost-zero-energy eigenstates are approximated given as
\begin{eqnarray}\label{Eigenstate} 
|0_{+}\rangle&=&\frac{|L\rangle+(-1)^{N+1}|R\rangle}{\sqrt{2}},~~
E_{+}=\Big|\frac{v\eta^{N-1}(\eta^{2}-1)}{\eta^{2N}-1}\Big|,\cr\cr
|0_{-}\rangle&=&\frac{|L\rangle-(-1)^{N+1}|R\rangle}{\sqrt{2}},~~
E_{-}=-\Big|\frac{v\eta^{N-1}(\eta^{2}-1)}{\eta^{2N}-1}\Big|.\nonumber\\
\end{eqnarray}
Here
\begin{eqnarray}\label{EdgeState} 
|L\rangle&=&|1\rangle+\eta|3\rangle+\cdots+\eta^{N-1}|2N-1\rangle,\cr\cr
|R\rangle&=&|2N\rangle+\eta|2N-2\rangle+\cdots+\eta^{N-1}|2\rangle,
\end{eqnarray}
denote the ideal exponentially localized left and right edge states in thermodynamic limit and $\eta=-v/w$ is the localization factor (See Appendix~\ref{appA} for more details). $|n\rangle=\hat{b}_{n}^{\dag}|V\rangle$ denote the state of the chain where the boson is on $n$-th site with $|V\rangle$ being vacuum state. The hybridized edge states $|0_{+}\rangle$ and $|0_{-}\rangle$ are superpositions of states localized exponentially on the left and right edge. What’s more, the odd or even superposition depends on the parity of total number $N$ of the unit cells.

The hybridized edge states $|0_{+}\rangle$ and $|0_{-}\rangle$ are also eigenstates of the inversion operator and hold opposite parity, 
\begin{eqnarray}\label{InversionOperator} 
\hat{I}=\sum_{n=1}^{2N}|2N+1-n\rangle\langle n|,~~~~~~~~~~~~~~\cr\cr
\hat{I}|0_{+}\rangle=(-1)^{N+1}|0_{+}\rangle,~~~~
\hat{I}|0_{-}\rangle=(-1)^{N}|0_{-}\rangle,
\end{eqnarray}
with $\hat{I}$ being real space form of inversion symmetry operator. And then the evolution of two hybridized edge states can be treated separately. What's more, inversion symmetry ensure that the distribution of eigenstates on left and right part of SSH chain are equal, that is 
\begin{eqnarray}\label{DF}
D_{f}\equiv\sum_{i=1}^{N}|c_{i}|^{2}-\sum_{j=N+1}^{2N}|c_{j}|^{2}=0.
\end{eqnarray}
Here $|c_{i}|^{2}$ is the distribution of $i$-th site (See Appendix~\ref{appB} for more details). 

%%%%%%%%%%%%%%%%%%%%%%%%%%%%%%%%%%%%%%%%%%%%%%%%%%%%%%%%%%%%%%%%%%%%%%%%%%%%%%%%%%%%
\section{TUNABLE BEAM SPLITTER AND HONG-OU-MANDEL INTERFERENCE VIA SSH CHAIN}\label{Sec3}
%%%%%%%%%%%%%%%%%%%%%%%%%%%%%%%%%%%%%%%%%%%%%%%%%%%%%%%%%%%%%%%%%%%%%%%%%%%%%%%%%%%%
%%%%%%%%%%%%%%%%%%%%%%%%%%%%%%%%%%%%%%%%%%%%%%%%%%%%%%%%%%%%%%%%%%%%%%%%%%%%%%%%%%%%
\subsection{Tunable beam splitter}\label{Sec31}
%%%%%%%%%%%%%%%%%%%%%%%%%%%%%%%%%%%%%%%%%%%%%%%%%%%%%%%%%%%%%%%%%%%%%%%%%%%%%%%%%%%%
\begin{figure}[!htp]
\centering
\includegraphics[width=0.8\columnwidth]{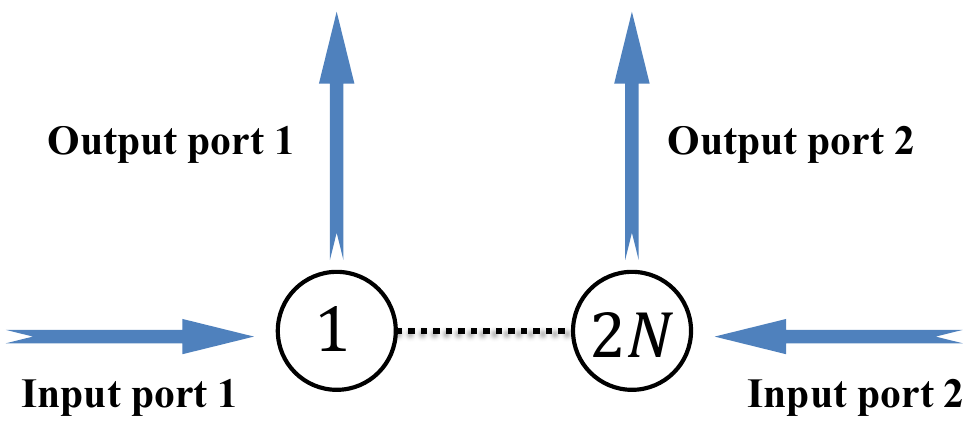}
\caption{\label{BSModel} Schematic diagram of the tunable BS. Site $1$ acts as port $1$ and site $2N$ acts as port $2$. A boson injected into the input port $1$ or $2$ can be transferred only to the output ports $1$ and $2$ with tunable proportion. From this perspective, the present system is equivalent to a tunable BS.}
\end{figure}

In this section, we explore how to achieve a tunable BS via edge channels in SSH model. We consider two end sites as both input ports and output ports, as shown in Fig.~\ref{BSModel}. A boson injected into site $1$ (site $2N$) can be transferred only to the two output ports with tunable proportion.

\begin{figure}[!htp]
\centering
\includegraphics[width=1\columnwidth]{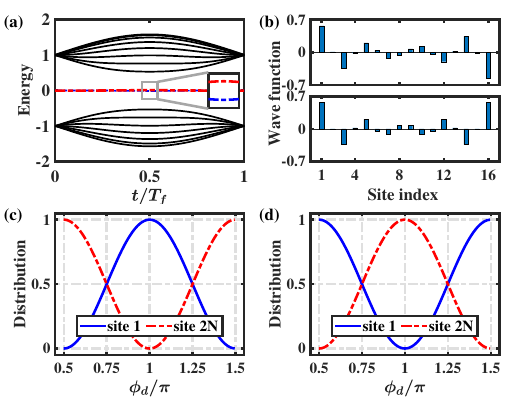}%
\caption{\label{Energy}(a) The energy spectrum for an finite sized SSH model with $2N=16$ and $v_{0}=0.6$. Solid black lines represent the bulk states and dashed color lines represent the hybridized edge states. (b) Wave functions of the hybridized edge states $|0_{+}\rangle$ (up) and $|0_{-}\rangle$ (down) at time $t=T_f/2$. (c), (d) The distribution of the final state at the output ports $1$ ($|1\rangle$,solid blue line) and $2$ ($|2N\rangle$, dashed red line) versus the dynamical phase $\phi_{d}$ with a boson initially injected in input ports $1$ and $2$, respectively. }
\end{figure}

We start by considering the topological fully dimerized limit ($v=0$). The edge state $|L\rangle$ and $|R\rangle$ localized only in site $1$ and $2N$ respectively according to Eq.~\ref{EdgeState}. A boson initially injected in site $1$ ($|1\rangle$) can be rewritten as the superpositions of two hybridized edge states $|0_{+}\rangle$ and $|0_{-}\rangle$, i.e., $|1\rangle=(|0_{+}\rangle+|0_{-}\rangle)/\sqrt{2}$. Then we move away from the fully dimerized limit by adiabatically changing the intracell hopping amplitudes and taking intercell ones as constant
\begin{eqnarray}\label{AdiabaticChange} 
v(t)=v_{0}\sin\theta(t).
\end{eqnarray}
Here $v_{0}<1$ is the modulation amplitude of the intracell hopping amplitudes, $\theta(t)=\pi t/T_{f}$ varying linearly with the adiabatic evolution time $t\in[0,T_{f}]$, and $T_{f}$ is the total evolution time in the units of $1⁄w$.
For a SSH chain with $2N=16$ sites, the energy spectrum versus evolution time $t$ is plotted in Fig.~\ref{Energy}(a) and the hybridized topological edge states $|0_{+}\rangle$ and $|0_{-}\rangle$ are well separated from the bulk states. At initial time $t=0$, the SSH chain is in the fully dimerized limit, edge states are $|L\rangle=|1\rangle$ and $|R\rangle=|2N\rangle$ edge states $|L\rangle=|1\rangle$ and $|R\rangle=|2N\rangle$ associating with eigenvalue $E_{+}=-E_{-}=0$. When it evolves to $\theta=\pi/2$, the SSH chain is moving away from the fully dimerized limit, $|E_{\pm}|$ gets the maximum value and the distribution of $|L\rangle$ ($|R\rangle$) is among different odd (even) sites [Fig.~\ref{Energy}(b)]. After then, the system returns to the fully dimerized limit at $\theta=\pi$.

As mentioned in Sec.~\ref{Sec2}, the eigenstates $|0_{+}\rangle$ and $|0_{-}\rangle$ will evolve separately in different sectors of the Hilbert space labeled by their parity. Following quantum adiabatic theorem, at time $t$ the initial state $|1\rangle$ should evolve to
\begin{eqnarray}\label{phit} 
|\psi(t)\rangle=\frac{e^{-i\int_{0}^{t}E_{+}d\tau}|0_{+}(t)\rangle
+e^{-i\int_{0}^{t}E_{-}d\tau}|0_{-}(t)\rangle}{\sqrt{2}}.
\end{eqnarray}
As a result, when the system returns to the fully dimerized ($t=T_{f}$) limit, the final state is given as
\begin{eqnarray}\label{phiTf} 
|\psi_{F}\rangle&=&\frac{e^{-i\phi_{d}}|0_{+}(T_{f})\rangle
+e^{i\phi_{d}}|0_{-}(T_{f})\rangle}{\sqrt{2}}\cr\cr
&=&\cos\phi_{d}|1\rangle+(-1)^{N}\sin\phi_{d}|2N\rangle,
\end{eqnarray}
where $\phi_{d}=\int_{0}^{T_{f}}E_{+}(t)dt=-\int_{0}^{T_{f}}E_{-}(t)dt$ is the dynamical phase accumulated in the adiabatic evolution. Here we use the symmetry of the energy spectrum i.e., $E_{+}=-E_{-}$. The dynamical phase depends on the total adiabatic evolution time and we can get desired $\phi_{d}$ by choosing suitable $T_{f}$. Obviously, the boson initially injects in the input port $1$ can be finally observed at the output ports $1$ and $2$ with a certain proportion. What's more, the splitting proportion can be arbitrarily tuned from $1$ to $0$ by modulating the dynamical phase $\phi_{d}$. If the boson is initially injecting in the input port $2$, we have the similar result, and the tunable BS is written as 
\begin{eqnarray}\label{TBS} 
|1\rangle
&\rightarrow&\cos\phi_{d}|1\rangle+(-1)^{N}i\sin\phi_{d}|2N\rangle,\cr\cr
|2N\rangle
&\rightarrow&(-1)^{N}i\sin\phi_{d}|1\rangle+\cos\phi_{d}|2N\rangle.
\end{eqnarray}
In conclusion, the realization of tunable BS has to satisfy two
key conditions: (i) the hybridized edge states $|0_{+}\rangle$ and $|0_{-}\rangle$ could evolve separately, and (ii) the dynamical phase $\phi_{d}$ is well controlled.

We plot the distribution of the final state at output ports $1$ ($|1\rangle$, solid blue line) and $2$ ($|2N\rangle$, dashed red line) vs the dynamical phase $\phi_{d}\in[\pi/2,3\pi/2]$ in Fig.~\ref{Energy}(c) and (d). These numerical results are in good agreement with analytical ones in Eq.~\ref{TBS}. In Fig.~\ref{Energy}(c) the boson is initially injected in input port $1$ ($|1\rangle$) and the final distribution in the output port $1$ can increase from $0$ to $1$ while the distribution in the output port $2$ decrease from $1$ to $0$ simultaneously at range $[\phi_{d}:\pi/2\rightarrow\pi]$. And their behaviour will be reversed for range $[\phi_{d}:\pi\rightarrow3\pi/2]$. Similar results will be obtained for the boson initially injected in input port $2$, which is shown in Fig.~\ref{Energy}(d).
 
Above phenomena illuminates we can achieve a tunable BS via edge channel in the SSH model. These two end sites of chain act as input and output ports. Above tunable BS has numerous potential applications in quantum optics and quantum information processing with different splitting proportions, such as optical trapping or photon storage ($100:0$ and $0:100$), quantum state transfer ($0:100$), photon distributions between two distant nodes (arbitrary splitting proportion), and HOM interference ($50:50$).
 
%%%%%%%%%%%%%%%%%%%%%%%%%%%%%%%%%%%%%%%%%%%%%%%%%%%%%%%%%%%%%%%%%%%%%%%%%%%%%%%%%%%%
\subsection{Hong-Ou-Mandel interference}\label{Sec32}
%%%%%%%%%%%%%%%%%%%%%%%%%%%%%%%%%%%%%%%%%%%%%%%%%%%%%%%%%%%%%%%%%%%%%%%%%%%%%%%%%%%%
In this section, we focus on two-boson HOM interference via the tunable BS introduced in Sec.~\ref{Sec31}. We begin evolution at $t=0$ and two identical bosons respectively injected in site $1$ and $2N$, i.e., the two-boson initial state is $|1,2N\rangle$. Then, we adiabatically tune the intracell hopping amplitudes according to Eq.~\ref{AdiabaticChange}. At the final time $t=T_{f}$, as illustrated in Sec.~\ref{Sec31}, the bosons will undergo tunable BS operation according to Eq.~\ref{TBS}. As a result, the two identical bosons initial in state $|1,2N\rangle$ will evolve to $(-1)^{N}i\sin2\phi_{d}(|1,1\rangle+|2N,2N\rangle)/\sqrt{2}+\cos2\phi_{d} |1,2N\rangle$. 
Intriguingly, when we take $\phi_{d}=(2l+1)\pi⁄4,l\in{Z}$, i.e., $50:50$ BS, the antibunching term will disappear. Thus, above state evolves into a two-particle NOON state
\begin{eqnarray}\label{phiNOON} 
|\psi_{{\rm NOON}}\rangle=\frac{1}{\sqrt{2}}(|1,1\rangle+|2N,2N\rangle),
\end{eqnarray}
where the global phase factor is ignored. The spatial NOON state is a maximally entangled state between two-end sites due to the well-known HOM interference.

\begin{figure}[!htp]
\centering
\includegraphics[width=1\columnwidth]{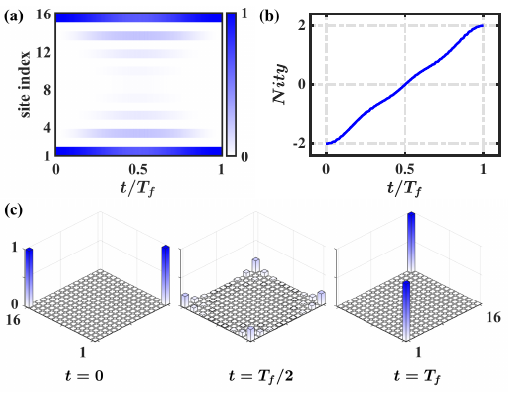}%
\caption{\label{HOM}(a), (b) Density distribution $\langle\hat{n}_{r}\rangle$ and $Nity$ as function of evolution time $t$. (c) Two-particle correlation $\Gamma_{q,r}$ at several typical evolution times. Here, total evolution time is $T_f=252$ in unit of $1/w$, associating dynamical phase accumulated in the adiabatic evolution will be $\phi_{d}=\pi/4$. And two-boson initial state is $|1,2N\rangle$. Other parameters are set as $2N=16$, $v_0=0.6$.}
\end{figure}

To explore correlation feature and characterize entanglement of the generated two-boson NOON state, we calculate density distribution $\langle\hat{n}_{r}\rangle$, two-particle correlation $\Gamma_{q,r}$ and ``NOONity'' ($Nity$),
\begin{eqnarray}\label{NOONity} 
\langle\hat{n}_{r}\rangle&=&\langle\hat{b}_{r}^{\dag}\hat{b}_{r}\rangle,\cr\cr
\Gamma_{q,r}&=&\langle\psi|\hat{b}_{q}^{\dag}\hat{b}_{r}^{\dag}\hat{b}_{r}\hat{b}_{q}|\psi\rangle,\cr\cr
Nity&=&\sum_{q,r}\Gamma_{q,q}\Gamma_{r,r}-\Gamma_{q,r}^{2}.
\end{eqnarray}
whose values during adiabatic evolution are given in Fig.~\ref{HOM}. 
Both the density distribution of the initial state and final state at $t=T_{f}$ just hold nonzero values ($1$) at $1$ and $2N$,  as shown in Fig.~\ref{HOM}(a), while they behave huge difference in correlation and entanglement features. Specifically, the value of $Nity$ is given in Fig.~\ref{HOM}(b), which must fall in the range $[-2,2]$ and will be larger if the state is more like the NOON state. Thereby, the increase of ``NOONity'' from $Nity=-2$ to $Nity=2$ confirms the forming of ideal two-particle NOON state from initial product state $|1,2N\rangle$. More intuitively, the correlations $\Gamma_{q,r}$ at several typical evolution times $t=0, T_{f}⁄2, T_{f}$ in Fig.~\ref{HOM}(c) display the process where correlation function changes from uncorrelated antibunching to correlated bunching patterns.

%%%%%%%%%%%%%%%%%%%%%%%%%%%%%%%%%%%%%%%%%%%%%%%%%%%%%%%%%%%%%%%%%%%%%%%%%%%%%%%%%%%%
\subsection{Disordered SSH chain in class BDI}\label{Sec33}
%%%%%%%%%%%%%%%%%%%%%%%%%%%%%%%%%%%%%%%%%%%%%%%%%%%%%%%%%%%%%%%%%%%%%%%%%%%%%%%%%%%%
After establishing tunable BS and HOM interference in SSH model with clean limit, we began introducing disorder and investing the role of distinct symmetry for protection above two phenomena.
In this section, we focus on the disordered chains remaining in BDI class, i.e., the time-reversal and chiral symmetry are not broken by disorder. The real hopping amplitude disorder conform to this requirement. As we can see in following, the chiral symmetry and time-reversal symmetry together could protect the tunable BS and HOM interference against disorder. More specifically, the parameters with disorder can be read as
\begin{eqnarray}\label{diorder1} 
v_{2n-1}(t)=v(t)(1+\xi{r_{2n-1}}),~w_{2n}=w(1+\xi{r_{2n}}),
\end{eqnarray}
where $\xi$ is the disorder strength and $r_{n}\in[-0.5,0.5]$ is a uniformly distributed random number. Generally speaking this disorder could break inversion symmetry and the eigenstates no long hold definite parity. However, we will see chiral symmetry ensures that there is no transition between eigenstate $|\psi_{n}\rangle$ and its chiral symmetric partner $\hat{S}|\psi_{n}\rangle$ during evolution. The derivative of disordered Hamiltonian $\mathcal{\hat{H}}$ also preserve chiral symmetry i.e., $\{d\mathcal{\hat{H}}/dt, \hat{S}\}=0$. Under time-reversal symmetry, $\langle\psi_{n}|d\mathcal{\hat{H}}/dt\hat{S}|\psi_{n}\rangle=\langle\psi_{n}|\hat{S}d\mathcal{\hat{H}}/dt|\psi_{n}\rangle$ are both real number and we also have $\langle\psi_{n}|d\mathcal{\hat{H}}/dt\hat{S}+\hat{S}d\mathcal{\hat{H}}/dt|\psi_{n}\rangle$=0. As a result, we have  
\begin{eqnarray}
\langle\psi_{n}|\frac{d\mathcal{\hat{H}}}{dt}\hat{S}|\psi_{n}\rangle
=\langle\psi_{n}|\hat{S}\frac{d\mathcal{\hat{H}}}{dt}|\psi_{n}\rangle
=0,
\end{eqnarray}
which means there is no transition between eigenstate $|\psi_{n}\rangle$ and $\hat{S}|\psi_{n}\rangle$. In this way the hybridized edge states $|0_{+}\rangle$ and $|0_{-}\rangle$ could evolve separately though inversion symmetry is broken.

There are two distinct treatments about the disorder, i.e., time independent static noise or time dependent temporal noise. When we consider static case, the disorder is fixed during the adiabatic evolution for each realization of the numerical simulation. In contrast, it is fluctuated during the evolution in temporal case. Here, we study both static and temporal cases. 
  
 \begin{figure}[!htp]
\centering
\includegraphics[width=1\columnwidth]{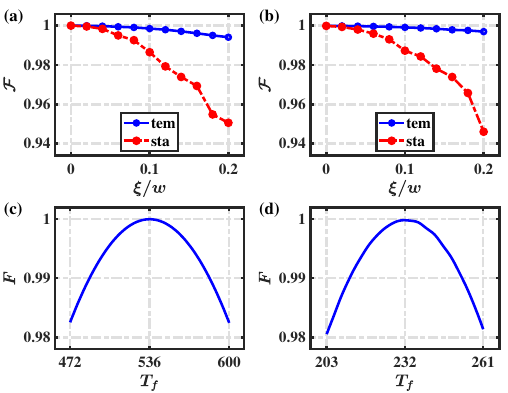}%
\caption{\label{Chiral}Protection of chiral symmetry. (a), (b) Average fidelities of tunable BS ($0:100$) and HOM interference, respectively, with different disorder strength. The initial state is $|1\rangle$, $T_f=504$ in unit of $1/w$, and $\phi_d=\pi/2$ same to tunable BS under clean limit; the initial state is $|1,2N\rangle$, $T_f=252$, and $\phi_d=\pi/4$ in HOM interference. (c), (d) Fidelity of tunable BS ($0:100$) and HOM interference, respectively, as function of total evolution time $T_f$, where disorder is static and takes strength $\xi=0.2$. Other parameters are set as $2N=16$, $v_0=0.6$.}
\end{figure}
 
For each realization we calculate the fidelity at a given final time $T_{f}$ and then take the average over $100$ repetitions of the adiabatic evolution. Each realization has its different random choice of disorder. In Fig.~\ref{Chiral}(a), we plot average values $\mathcal{F}$ of fidelity $F=|\langle\psi_{F}|\psi(T_{f})\rangle|$ for different disorder strength $\xi$. 
The initial state is also set as $|1\rangle$, i.e., a boson is injected in input port $1$ at initial time $t=0$ and we take $T_f=504$ in unit of $1/w$ and $\phi_d=\pi/2$ correspond to $0:100$ BS. With this setup, the boson injected in input port $1$ ($2$) can be transferred to output port $2$ ($1$) under clean limit. In current disordered cases as discussed in Sec.~\ref{Sec31}, we find that in temporal case (solid blue line) there is a plateau near $1$ for $\xi$ in the range $\xi/w\in[0,0.2]$. However, it will decrease from $1$ to $0.94$ in static case (dashed red line). The results of HOM interference against this disorder is similar, see average values $\mathcal{F}$ of fidelity $F=|\langle\psi_{{\rm NOON}}|\psi(T_{f})\rangle|$ in Fig.~\ref{Chiral}(b). 

The difference of static and temporal disorder comes from the influence of dynamical phase $\phi_d$. More specifically, the chiral-symmetry-preserving hopping amplitude disorder only make a tiny random shift on the energy spectrum and influence the dynamical phase. In the temporal case with time dependent disorder, the influence on the dynamical phase can be averaged out during the adiabatic evolution. Therefore, for our protocol, the average fidelity
$\mathcal{F}$ will remain above $0.99$ as long as disorder strength $\xi$ is less than $0.2w$ in temporal case and the total evolution time $T_f$ are same as the clean case [Fig.~\ref{Chiral}(a) and (b)]. In the static case, the disorder fixed during the adiabatic evolution, we should adjust the total evolution time $T_f$ to match the desired dynamical phase [Fig.~\ref{Chiral}(c) and (d)]. 

%%%%%%%%%%%%%%%%%%%%%%%%%%%%%%%%%%%%%%%%%%%%%%%%%%%%%%%%%%%%%%%%%%%%%%%%%%%%%%%%%%%%%%%%%%%%%%%%%%%%%%%%%%%%%%%%%%%%%%%%%%%%%%%%%%%%%%%
\section{Exact inversion symmetry and temporal-average-inversion-symmetry}\label{Sec4}
%%%%%%%%%%%%%%%%%%%%%%%%%%%%%%%%%%%%%%%%%%%%%%%%%%%%%%%%%%%%%%%%%%%%%%%%%%%%%%%%%%%%
In Sec.~\ref{Sec33}, we have investigated hopping amplitude disorder (preserving chiral symmetry but breaking inversion symmetry) and illustrated the protection of chiral symmetry and time-reversal symmetry. However chiral symmetry may easily broken by several common disorders such as on-site disorder. It is necessary to consider what happens when chiral symmetry is broken, but inversion symmetry preserves. In this section, we will characterize the protection of inversion symmetry both in exact case and temporal-averaged one.

%%%%%%%%%%%%%%%%%%%%%%%%%%%%%%%%%%%%%%%%%%%%%%%%%%%%%%%%%%%%%%%%%%%%%%%%%%%%%%%%%%%%%%%%%%%%%%%%%%%%%%%%%%%%%%%%%%%%%%%%%%%%%%%%%%%%%%%
\subsection{Static disorder preserving or breaking inversion symmetry}\label{Sec41}
%%%%%%%%%%%%%%%%%%%%%%%%%%%%%%%%%%%%%%%%%%%%%%%%%%%%%%%%%%%%%%%%%%%%%%%%%%%%%%%%%%%%
We began with the static on-site disorder described by 
\begin{eqnarray}\label{diorder2} 
\hat{H}_{\zeta}=\zeta\sum_{n}r_{n}\hat{b}_{n}^{\dag}\hat{b}_{n},
\end{eqnarray}
where $\zeta$ is the disorder strength and $r_{n}\in[-0.5,0.5]$ is a uniformly distributed random number. In general, this disorder breaks both chiral and inversion symmetry. To preserve inversion symmetry, the disorder configuration
must respect to the inversion center, that is the random number $r_{n}$ should satisfy 
\begin{eqnarray} 
r_{n}=r_{2N+1-n}.
\end{eqnarray}
We will consider both inversion-symmetry-preserving on-sit disorder and generic on-site disorder (breaking inversion symmetry) in this subsection.

\begin{figure}[!htp]
\centering
\includegraphics[width=1\columnwidth]{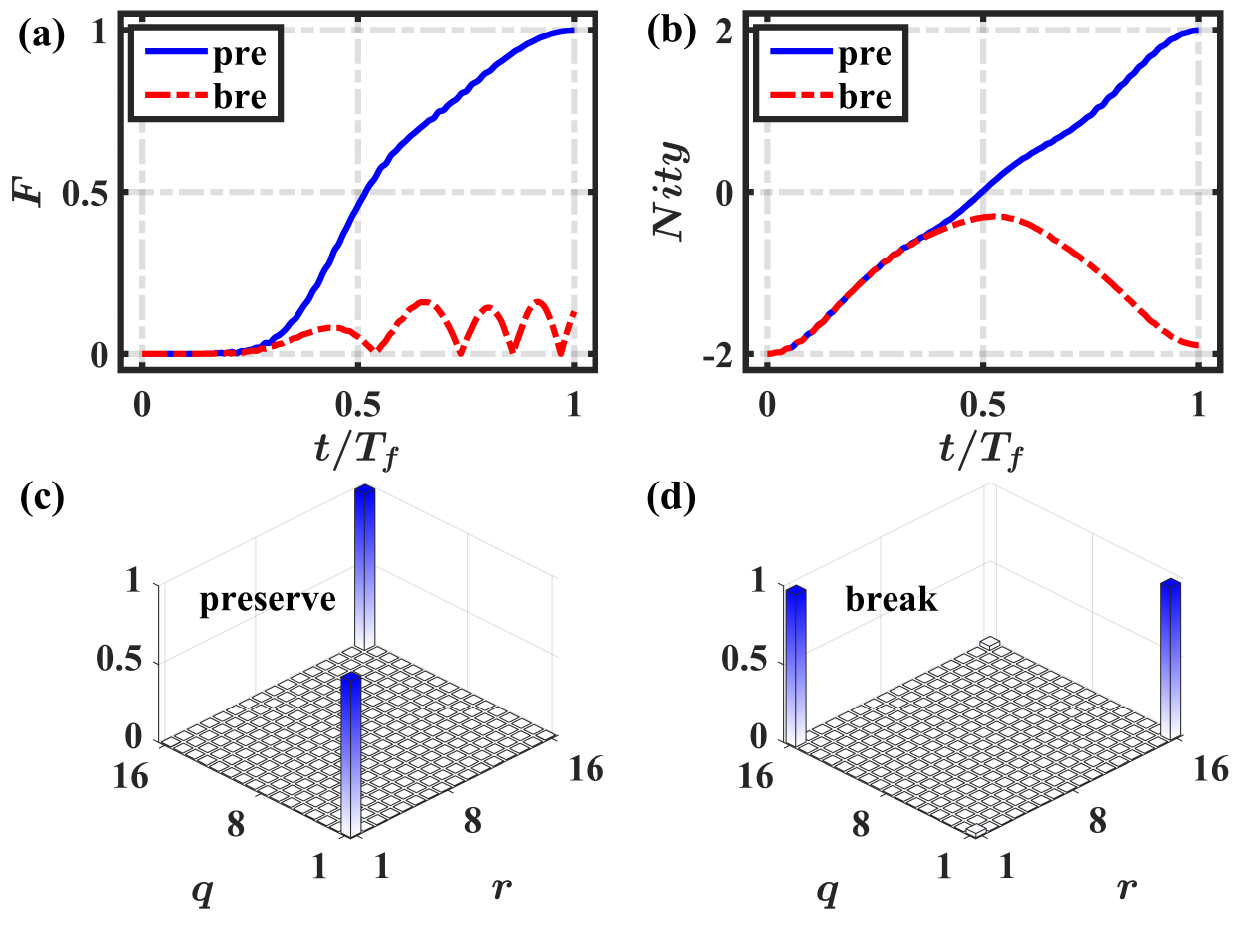}%
\caption{\label{Inversion}Protection of exact inversion symmetry. (a), (b) Fidelity $F$ and $Nity$ during the HOM interference procession, respectively. (c), (d) Two-particle correlation $\Gamma_{q,r}$ at $t=T_f$ for cases with inversion-symmetry-preserving and -breaking disorder, respectively. The two-particle input state is $|1,2N\rangle$ and the ideal NOON state generate by HOM interference is $
|\psi_{{\rm NOON}}\rangle=(|1,1\rangle+|2N,2N\rangle)/\sqrt{2}$. The disorder strength $\zeta=0.2w$ and other parameters are set as $2N=16, T_f=252$.}
\end{figure}

To illustrate the protection of inversion symmetry to HOM interference, we also calculate the fidelity $F=|\langle\psi_{{\rm NOON}}|\psi(t)\rangle|$, $Nity$, and two-particle correlation $\Gamma_{q,r}$ in the interference procession. 
In Fig.~\ref{Inversion}(a), we plot the fidelity for the input state $|1,2N\rangle$ as a function of evolution time $t$ during HOM interference with disorder strength $\zeta=0.2$. Under inversion-symmetry-preserving disorder (blue sold line), the fidelity $F=0$ for the input product state will increase to $F=1$ at $t=T_f$. Indeed we generate a NOON state by HOM interference and take the total evolution time $T_f=252$ as the same to clean situation. In contrast, the generic on-site disorder could break inversion symmetry and the fidelity (red dashed line) always close to $0$. To explore the entanglement and correlation feature, we also provide $Nity$ and $\Gamma_{q,r}$ in Figs.~\ref{Inversion}(b)-(d). As shown in Fig.~\ref{Inversion}(b), $Nity=-2$ for the input state $|1,2N\rangle$ and it increase to $2$ at final evolution time $t=T_f$ under inversion-symmetry-preserving disorder (blue sold line). In contrast, it return to $-2$ at $t=T_f$ under inversion-symmetry-breaking case (red dashed line). In Figs.~\ref{Inversion}(c) and (d), We present two-particle correlation $\Gamma_{q,r}$ at $t=T_f$. The final correlation function still take uncorrelated antibunching pattern under inversion-symmetry-breaking case [Fig.~\ref{Inversion}(d)] while behaves correlated bunching pattern for the case preserving inversion symmetry [Fig.~\ref{Inversion}(c)], consistent with the evolution of fidelity $F$ and $Nity$ in Figs.~\ref{Inversion}(a) and (b). Therefore, exact inversion symmetry could protect the HOM interference even though chiral symmetry is breaking, which can be further confirmed by investing the conservation of parity during evolution process as given in Appendix~\ref{appB}.
 
%%%%%%%%%%%%%%%%%%%%%%%%%%%%%%%%%%%%%%%%%%%%%%%%%%%%%%%%%%%%%%%%%%%%%%%%%%%%%%%%%%%%%%%%%%%%%%%%%%%%%%%%%%%%%%%%%%%%%%%%%%%%%%%%%%%%%%%
\subsection{Emergence of temporal-average-inversion-symmetry}\label{Sec42}
%%%%%%%%%%%%%%%%%%%%%%%%%%%%%%%%%%%%%%%%%%%%%%%%%%%%%%%%%%%%%%%%%%%%%%%%%%%%%%%%%%%%
Although we can preserve inversion symmetry by engineering disorder symmetrically~\cite{PhysRevB.103.024205}, it's more important to restore inversion symmetry in a generic way. Previous works~\cite{PhysRevB.86.115112,PhysRevB.87.035119,LFuPRL2012,PhysRevB.89.155424,PhysRevResearch.2.012067,PhysRevLett.126.206404,PhysRevX.13.031016,Tao2023} have demonstrated an average spatial symmetry on the statistical ensemble of disordered system can protect topological phase. In this section, we consider temporal case of generic on-site disorder, where temporal-average-inversion-symmetry emerges. Specially, disorder descried in Eq.~\ref{diorder2} will fluctuate during the evolution for each realization of numerical simulation.

\begin{figure}[!htp]
\centering
\includegraphics[width=1\columnwidth]{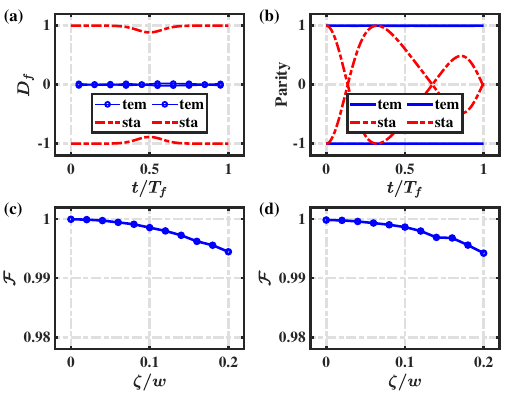}%
\caption{\label{Parity1}Protection of temporal-average-inversion-symmetry. (a) Distribution difference of the two in-gap instantaneous eigenstates in static (dashed red line) and temporal (solid blue line) case, respectively. (b) Parity during adiabatic evolution in static (dashed red line) and temporal (solid blue line) case, respectively. The initial state are $(|1\rangle+|2N\rangle)/\sqrt{2}$ or $(|1\rangle-|2N\rangle)/\sqrt{2}$. The total evolution time in (a) and (b) are $T_f=252$. (c), (d) Average fidelities of tunable BS ($0:100$) and HOM interference with different disorder strength, respectively, for temporal case. The initial state is $|1\rangle$, $T_f=504$ in unit of $1/w$, and $\phi_d=\pi/2$ same to the clean case in tunable BS ($0:100$); the initial state is $|1,2N\rangle$, $T_f=252$, and $\phi_d=\pi/4$ in HOM interference. Other parameters are set as $2N=16$, $v_0=0.6$.}
\end{figure}

As discussed in Sec.~\ref{Sec41}, our protocol is destroyed by static on-site disorder breaking inversion symmetr6y. To explore the difference between static and temporal case, we calculate distribution difference $D_{f}$ defined in Eq.~\eqref{DF} of in-gap instantaneous eigenstates. As shown in Fig,~\ref{Parity1}(a), the average distribution difference of in-gap instantaneous eigenstates are closed to $0$ and each point is average about a duration of $0.1T_f$. The result is agree with clean limit and inversion-symmetry-preserving static on-sit disorder given in Appendix~\ref{appB}. In contrast it close to $1$ or $-1$ in static case and the in-gap instantaneous eigenstates are long hybridized edge states. Associating with $D_{f}\approx0$, the parity $P=\langle\hat{I}\rangle$ during adiabatic evolution is conserved during adiabatic evolution in temporal case [Fig,~\ref{Parity1}(b)]. In conclusion, there is temporal-average-inversion-symmetry in temporal case, restoring the parity conservation existing in exact inversion symmetric cases. Protected by this temporal-average-inversion-symmetry, hybridized edge states $|0_{+}\rangle$ and $|0_{-}\rangle$ will evolve separately, and then tunable BS and HOM interference could also be restored, as shown in Figs.~\ref{Parity1}(c) and (d). Here the average fidelity $\mathcal{F}$ will remain above $0.99$ as long as disorder strength $\xi$ is less than $0.2w$ in temporal case and the total evolution time $T_f$ are as the same in clean limit. 

%%%%%%%%%%%%%%%%%%%%%%%%%%%%%%%%%%%%%%%%%%%%%%%%%%%%%%%%%%%%%%%%%%%%%%%%%%%%%%%%%%%%%%%%%%%%%%%%%%%%%%%%%%%%%%%%%%%%%%%%%%%%%%%%%%%%%%%
\section{SUMMARY AND DISCUSSION}\label{Sec5}
%%%%%%%%%%%%%%%%%%%%%%%%%%%%%%%%%%%%%%%%%%%%%%%%%%%%%%%%%%%%%%%%%%%%%%%%%%%%%%%%%%%%
Based upon the nontrivial boundary states of topological phases, we have established tunable BS and HOM interference in SSH chain, and elucidated effects of distinct symmetries.
To realize these behaviours, the hybridized edge states should evolve separately and accumulate dynamical phase difference.
We reveal that chiral and time-reversal symmetries could protect this process via adiabatic theorem and examine it numerically by introducing associating disorders. 
For more practical applications, we turn to study the role of inversion symmetry and find the parity conservation protected by this symmetry also can ensure this mechanism.
Despite various type of disorders could break exact inversion symmetry and destroy BS as well as HOM interference, a temporal-average-inversion-symmetry appears when disorder becomes temporal.
This emergent symmetry also can protect parity conservation, giving rise to temporal-average-inversion-symmetry protected tunable BS and HOM interference. 
Obviously, our approach can be applied to a large number of realistic systems, considering the widely existence of temporal disorders. 
Moreover, we point out that our scheme can be extended to interacting topological states~\cite{YKePRA2017,WLiuPRR2023,WalterNP2023}.

Lastly, we briefly discuss the experimental feasibility of our protocol. 
Photonic waveguide arrays represent a promising platform for exploring topological features and several topological effects have been observed using this system including non-quantized adiabatic pumping of topological edge state~\cite{YEKrausPRL2012}, topological protection of correlation and entanglement~\cite{ABlancoScience2018,MWangNanophotonics2019}, quantum interference of topological states of light~\cite{JLTambascoSA2018},and 4D quantum Hall physics~\cite{ZilberbergNature2018}.
The adiabatically modulated Aubry-Andr\'{e}-Harper model can be achieved by slowly varying the spacing between the waveguides along the propagation axis~\cite{YEKrausPRL2012,JLTambascoSA2018}.
One may use the adiabatically modulated photonic waveguide arrays as a potential platform for testing our protocols.

\acknowledgements{This work is supported by the National Natural Science Foundation of China under Grant No. 12104103, the Guangdong Basic and Applied Basic Research Foundation under Grant No. 2022A1515010726, and Science and Technology Program of Guangzhou under Grant No. 2023A04J0039.}

\appendix
\renewcommand\theequation{\Alph{section}\arabic{equation}}

%%%%%%%%%%%%%%%%%%%%%%%%%%%%%%%%%%%%%%%%%%%%%%%%%%%%%%%%%%%%%%%%%%%%%%%%%%%%%%%%%%%%
\section{Spectrum and edge sates constrained by chiral symmetry}\label{appA}
%%%%%%%%%%%%%%%%%%%%%%%%%%%%%%%%%%%%%%%%%%%%%%%%%%%%%%%%%%%%%%%%%%%%%%%%%%%%%%%%%%%%
In this section, we will discuss several properties of spectrum and eigenstates embodied in chiral symmetric systems.
Referring to the chiral symmetry, the single-particle
Hamiltonian of SSH chain must satisfy a anticommutation
relation
\begin{eqnarray}
\hat{S}\hat{H}{S}^{\dag}=-{H},
\end{eqnarray}
and the chiral symmetry operator $\hat{S}$ is unitary and Hermitian, i.e. $\hat{S}\hat{S}^{\dag}=\hat{S}^{2}=1$. The energy spectrum constrained by chiral symmetry is symmetric. For an eigenstate $|\psi_{n}\rangle$ with eigenenergy $E_n$, there is a chiral symmetric partner $\hat{S}|\psi_{n}\rangle$ with eigenenergy $-E_n$. This result with the minus sign is from the chiral symmetric relationship
\begin{eqnarray}
&\hat{H}|\psi_{n}\rangle=E_n|\psi_{n}\rangle\cr\cr
&\Rightarrow
\hat{H}\hat{S}|\psi_{n}\rangle=-\hat{S}\hat{H}|\psi_{n}\rangle
=-E_n\hat{S}|\psi_{n}\rangle,
\end{eqnarray}
For $E_n\neq0$, the states $|\psi_{n}\rangle$ and $\hat{S}|\psi_{n}\rangle$ are eigenstates with opposite energy and must be orthogonal. What's more, every nonzero energy eigenstate has equal support on both odd and even sites, which can be simply seen by
\begin{eqnarray}\label{EqualSupport}
0=\langle\psi_{n}|\hat{S}|\psi_{n}\rangle
=\langle\psi_{n}|\hat{P}_{{\rm odd}}|\psi_{n}\rangle
-\langle\psi_{n}|\hat{P}_{{\rm even}}|\psi_{n}\rangle.
\end{eqnarray}

It is well known that, in topological nontrivial phase ($|v|<|w|$), the SSH chain hosts two zero-energy edge states localized on the boundaries in the thermodynamic limit of $N\rightarrow\infty$. The edge state exponentially localized in the left and right sides of the chain are (un-normalized)
\begin{eqnarray}\label{EdgeStateA} 
|L\rangle&=&|1\rangle+\eta|3\rangle+\cdots+\eta^{N-1}|2N-1\rangle,\cr\cr
|R\rangle&=&|2N\rangle+\eta|2N-2\rangle+\cdots+\eta^{N-1}|2\rangle,
\end{eqnarray}
where $\eta=-v/w$ denotes the localization factor. We can derive the zero-energy edge state by solving $\hat{H}|L\rangle=0$ ($\hat{H}|R\rangle=0$) under left (right) semi-infinite boundary condition. The obtained right edge state $|R\rangle$ has nonvanishing components only on the even sites while the left edge state $|L\rangle$ distributes on the odd sites. In finite-sized system, the energies of the edge states remain very close to zero energy and take a pair of almost-zero-energy eigenvalues opposite to each other due to chiral symmetry [Fig.~\ref{Energy}(a) of main text]. 

Obviously, the left edge state $|L\rangle$ and the right edge state $|R\rangle$, localized only on the odd sites or even sites respectively, are no longer eigenstates according to Eq~\ref{EqualSupport} when $v\neq0$. Fortunately, we can obtain the almost-zero-energy eigenstates to a good approximation by diagonalizing the Hamiltonian under base vectors $|L\rangle$ and $|R\rangle$
\begin{eqnarray}
\langle{L}|\hat{H}|L\rangle&=&\langle{R}|\hat{H}|R\rangle=0,\cr\cr
\langle{L}|\hat{H}|R\rangle&=&\langle{R}|\hat{H}|L\rangle
=\frac{v\eta^{N-1}(\eta^{2}-1)}{\eta^{2N}-1}.
\end{eqnarray}
These almost-zero-energy eigenstates $|0_{+}\rangle$ and $|0_{-}\rangle$ can be obtained by
\begin{eqnarray}
|0_{+}\rangle&=&\frac{|L\rangle+(-1)^{N+1}|R\rangle}{\sqrt{2}},~~
E_{+}=\Big|\frac{\nu\eta^{N-1}(\eta^{2}-1)}{\eta^{2N}-1}\Big|,\cr\cr
|0_{-}\rangle&=&\frac{|L\rangle-(-1)^{N+1}|R\rangle}{\sqrt{2}},~~
E_{-}=-\Big|\frac{\nu\eta^{N-1}(\eta^{2}-1)}{\eta^{2N}-1}\Big|.\nonumber\\
\end{eqnarray}

%%%%%%%%%%%%%%%%%%%%%%%%%%%%%%%%%%%%%%%%%%%%%%%%%%%%%%%%%%%%%%%%%%%%%%%%%%%%%%%%%%%%
\section{Parity and inversion symmetry}\label{appB}
%%%%%%%%%%%%%%%%%%%%%%%%%%%%%%%%%%%%%%%%%%%%%%%%%%%%%%%%%%%%%%%%%%%%%%%%%%%%%%%%%%%%
Despite several internal symmetries, SSH chain also holds inversion symmetry $[\hat{H},\hat{I}]=0$.
Also, inversion symmetry operator $\hat{I}$ is unitary and Hermitian, i.e. $\hat{I}\hat{I}^{\dag}=\hat{I}^{2}=1$. As we all know, the eigenstate $|\psi_{n}\rangle$ of the system has certain parity under inversion symmetry  
\begin{eqnarray}
\hat{I}|\psi_{n}\rangle=\pm|\psi_{n}\rangle.
\end{eqnarray}
We can express $|\psi_{n}\rangle$ as
\begin{eqnarray}
|\psi_{n}\rangle=\sum_{i=1}^{2N}c_{i}|i\rangle,
\end{eqnarray}
and it satisfies $|c_{i}|=|c_{2N+1-i}|$. As a result, the distribution of eigenstates on left and right part of SSH chain are equal, that is 
\begin{eqnarray}\label{Df2}
\sum_{i=1}^{N}|c_{i}|^{2}=\sum_{j=N+1}^{2N}|c_{j}|^{2}.
\end{eqnarray}

\begin{figure}[!htp]
\centering
\includegraphics[width=1\columnwidth]{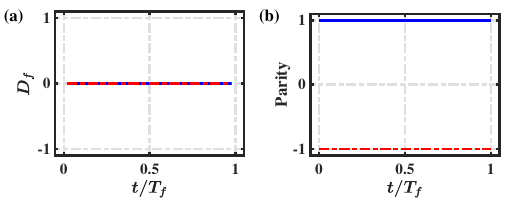}%
\caption{\label{Parity2}(a) Distribution difference of the two in-gap instantaneous eigenstates. (b) Conservation of parity during adiabatic evolution. The initial state are $(|1\rangle+|2N\rangle)/\sqrt{2}$ (solid blue line) and $(|1\rangle-|2N\rangle)/\sqrt{2}$ (dashed red line) , respectively. Other parameters are set as $2N=16,v_{0}=0.6, T_f=252$.}
\end{figure}

We now turn to consider the inversion-symmetry-preserving on-sit disorder introduced in Sec.~\ref{Sec41} of main text. 
To proceed, we calculate the distribution difference 
\begin{eqnarray}
D_{f}=\sum_{i=1}^{N}|c_{i}|^{2}-\sum_{j=N+1}^{2N}|c_{j}|^{2}.
\end{eqnarray}
of the two in-gap instantaneous eigenstates. As shown in Fig.~\ref{Parity2}(a), the distribution difference $D_{f}$ are exactly zero and the numerical result is well agree with Eq.~\ref{Df2}. Furthermore, we also calculate the parity during the evolution
\begin{eqnarray}
P=\langle\hat{I}\rangle,
\end{eqnarray}
with intimal stat $(|1\rangle+|2N\rangle)/\sqrt{2}$ and $(|1\rangle-|2N\rangle)/\sqrt{2}$ respectively. We present the numerical result in Fig.~\ref{Parity2}(b) and it shows conservation of parity clearly.


\begin{thebibliography}{99}
\bibitem{DXiaoRMP2010}
D. Xiao, M.-C. Chang, and Q. Niu, Berry phase effects on electronic properties, Rev. Mod. Phys. \textbf{82}, 1959 (2010).
\bibitem{MZHasanRMP2010}
M. Z. Hasan and C. L. Kane, Colloquium: Topological insulators, Rev. Mod. Phys. \textbf{82}, 3045 (2010).
\bibitem{XLQiRMP2011}
X.-L. Qi and S.-C. Zhang, Topological insulators and superconductors, Rev. Mod. Phys. \textbf{83}, 1057 (2011).

\bibitem{HaldanePRL2008}
F. D. M. Haldane and S. Raghu, Possible realization of directional optical waveguides in photonic crystals with broken time-reversal symmetry, Phys. Rev. Lett. \textbf{100},
013904 (2008).
\bibitem{ZWangNature2009}
Z. Wang, Y. Chong, J. D. Joannopoulos, and M. Soljačić, Observation of unidirectional backscattering-immune topological electromagnetic states. Nature \textbf{461}, 772 (2009).
\bibitem{LuNP2014}
L. Lu, J. D. Joannopoulos, and M. Soljačić, Topological photonics, Nat. Photonics \textbf{8}, 821 (2014).
\bibitem{TOzawaRMP2019}
T. Ozawa, H. M. Price, A. Amo, N. Goldman, M. Hafezi, L. Lu, M. C. Rechtsman, D. Schuster, J. Simon, O. Zilberberg, and I. Carusotto, Topological photonics, Rev. Mod. Phys. \textbf{91}, 015006 (2019).
%

\bibitem{ZYangPRL2015}
Z. Yang, F. Gao, X. Shi, X. Lin, Z. Gao, Y. Chong, and B. Zhang, Topological Acoustics, Phys. Rev. Lett. \textbf{114}, 114301 (2015).
\bibitem{MaNRP2019}
G. Ma, M. Xiao, and C. T. Chan, Topological phases in acoustic and mechanical systems, Nat. Rev. Phys. \textbf{1}, 281
(2019).
\bibitem{XueNRM2022}
H. Xue, Y. Yang, and B. Zhang, Topological acoustics, Nat. Rev. Mater. \textbf{7}, 974 (2022).



\bibitem{MAidelsburgerPRL2013}
M. Aidelsburger, M. Atala, M. Lohse, J. T. Barreiro, B. Paredes, and I. Bloch, Realization of the Hofstadter Hamiltonian with Ultracold Atoms in Optical Lattices, Phys. Rev. Lett. \textbf{111}, 185301 (2013).
\bibitem{CooperRMP2019}
N. R. Cooper, J. Dalibard, and I. B. Spielman, Topological bands for ultracold atoms, Rev. Mod. Phys. \textbf{91}, 015005 (2019).


\bibitem{YHatsugaiPRL1993}
Y. Hatsugai, Chern number and edge states in the integer quantum Hall effect, Phys. Rev. Lett. \textbf{71}, 3697 (1993).
\bibitem{EssinPRB2011}
A. M. Essin and V. Gurarie, Bulk-boundary correspondence of topological insulators from their respective green’s functions, Phys. Rev. B \textbf{84}, 125132 (2011).


\bibitem{Ryu_2010}
S. Ryu, A. P. Schnyder, A. Furusaki, and A. W. W. Ludwig, Topological insulators and superconductors: tenfold way and dimensional hierarchy, New J. Phys. \textbf{12}, 065010 (2010).
\bibitem{CKChiuRMP2016}
C.-K. Chiu, J. C. Y. Teo, A. P. Schnyder, and S. Ryu, Classification of topological quantum matter with symmetries, Rev. Mod. Phys. \textbf{88}, 035005 (2016).

\bibitem{JSeoNature2010}
J. Seo, P. Roushan, H. Beidenkopf, Y. S. Hor, R. J. Cava, and A. Yazdani, Transmission of topological surface states through surface barriers, Nature \textbf{466}, 343 (2010).

\bibitem{ABlancoScience2018}
A. Blanco-Redondo, B. Bell, D. Oren, B. J. Eggleton, and M. Segev, Topological protection of biphoton states, Science \textbf{362}, 568 (2018).

\bibitem{YEKrausPRL2012}
Y. E. Kraus, Y. Lahini, Z. Ringel, M. Verbin, and O. Zilberberg, Topological States and Adiabatic Pumping in Quasicrystals, Phys. Rev. Lett. \textbf{109}, 106402 (2012).
\bibitem{NLangQI2017}
N. Lang and H. P. Büchler, Topological networks for quantum communication between distant qubits, npj Quantum Inf. \textbf{3}, 47 (2017).
\bibitem{CDlaskaQST2017}
C. Dlaska, B. Vermersch, and P. Zoller, Robust quantum state transfer via topologically protected edge channels in dipolar arrays, Quantum Sci. Technol. \textbf{2}, 015001 (2017).
\bibitem{FMeiPRA2018}
F. Mei, G. Chen, L. Tian, S. L. Zhu, and S. Jia, Robust quantum state transfer via topological edge states in superconducting qubit chains, Phys. Rev. A \textbf{98}, 012331 (2018).
\bibitem{SLonghiPRB2019}
S. Longhi, Topological pumping of edge states via adiabatic passage, Phys. Rev. B \textbf{99}, 155150 (2019).
\bibitem{NEPalaiodimopoulosPRA2021}
N. E. Palaiodimopoulos, I. Brouzos, F. K. Diakonos, and G.Theocharis, Fast and robust quantum state transfer via a topological chain, Phys. Rev. A \textbf{103}, 052409 (2021).
\bibitem{LHuangPRA2022}
L. Huang, Z. Tan, H. Zhong, and B. Zhu, Fast and robust quantum state transfer assisted by zero-energy interface states in a splicing Su-Schrieffer-Heeger chain, Phys. Rev. A \textbf{106}, 022419 (2022).
\bibitem{CWangPRA2022}
C. Wang, L. Li, J. Gong, and Y.-X. Liu, Arbitrary entangled state transfer via a topological qubit chain, Phys. Rev. A \textbf{106}, 052411 (2022).

\bibitem{PBorossPRB2019}
P. Boross, J. K. Asbóth, G. Széchenyi, L. Oroszlány, and A. Pályi, Poor man’s topological quantum gate based on the Su-Schrieffer-Heeger model, Phys. Rev. B \textbf{100}, 045414 (2019).
\bibitem{MNarozniakPRB2021}
M. Naro\.{z}niak, M. C. Dartiailh, J. P. Dowling, J. Shabani, and T. Byrnes, Quantum gates for Majoranas zero modes in topological superconductors in one-dimensional geometry, Phys. Rev. B
\textbf{103}, 205429 (2021).




\bibitem{RHammerPRB2013}
R. Hammer and W. P\"{o}tz, Dynamics of domain-wall Dirac fermions on a topological insulator: A chiral fermion beam splitter, Phys. Rev. B \textbf{88}, 235119 (2013).
\bibitem{XSWangPRB2017}
X. S. Wang, Y. Su, and X. R. Wang, Topologically protected unidirectional edge spin waves and beam splitter, Phys. Rev. B \textbf{95}, 014435 (2017).
\bibitem{LQiPRB2021}
L. Qi, Y. Xing, X. D. Zhao, S. Liu, S. Zhang, S. Hu, and H. F. Wang, Topological beam splitter via defect-induced edge channel in the Rice-Mele model, Phys. Rev. B \textbf{103}, 085129 (2021).
\bibitem{LQiQuantum2021}
L. Qi, Y. Yan, Y. Xing, X. D. Zhao, S. Liu, X. Han, W. X. Cui, S. Zhang, and H. F. Wang, Tunable Topological Beam Splitter in Superconducting Circuit Lattice, Quantum Rep. \textbf{3}, 1 (2021).
\bibitem{LQiPRA2023}
L. Qi, N. Han, S. Hu, and A.-L. He, Engineering the unidirectional topological excitation transmission and topological diode in the Rice-Mele model, Phys. Rev. A \textbf{108}, 032402 (2023).

\bibitem{PhysRevB.55.1142}
A. Altland and M. R. Zirnbauer, Nonstandard symmetry classes in mesoscopic normal-superconducting hybrid structures, Phys. Rev. B \textbf{55}, 1142 (1997).

\bibitem{PhysRevLett.98.106803}
L. Fu, C. L. Kane, and E. J. Mele, Topological insulators in three dimensions, Phys. Rev. Lett. \textbf{98}, 106803 (2007).
\bibitem{Slager2013}
R.-J. Slager, A. Mesaros, V. Juri\v{c}i\'{c}, and J. Zaanen, The space group classification of topological band-insulators, Nat. Phys. \textbf{9}, 98 (2013).
\bibitem{PhysRevX.7.041069}
J. Kruthoff, J. de Boer, J. van Wezel, C. L. Kane, and R.-J. Slager, Topological classification of crystalline insulators through band structure combinatorics, Phys. Rev. X \textbf{7}, 041069 (2017).
\bibitem{Bradlyn2017}
B. Bradlyn, L. Elcoro, J. Cano, M. G. Vergniory, Z. Wang, C. Felser, M. I. Aroyo, and B. A. Bernevig, Topological quantum chemistry, Nature (London) \textbf{547}, 298 (2017).
\bibitem{Po2017}
H. C. Po, A. Vishwanath, and H. Watanabe, Symmetrybased indicators of band topology in the 230 space groups, Nat. Commun. \textbf{8}, 50 (2017).
\bibitem{Elcoro2021}
L. Elcoro, B. J. Wieder, Z. Song, Y. Xu, B. Bradlyn, and
B. A. Bernevig, Magnetic topological quantum chemistry,
Nat. Commun. \textbf{8}, 5965 (2021).

\bibitem{AEkertRMP1996}
A. Ekert and R. Jozsa, Quantum computation and Shor’s factoring algorithm, Rev. Mod. Phys. \textbf{68}, 733 (1996).
\bibitem{JWPanRMP2012}
J.-W. Pan, Z.-B. Chen, C.-Y. Lu, H. Weinfurter, A. Zeilinger, and M. Zukowski, Multiphoton entanglement and interferometry, Rev. Mod. Phys. \textbf{84}, 777 (2012).
\bibitem{LPezzeRMP2018}
L. Pezz\`{e}, A. Smerzi, M. K. Oberthaler, R. Schmied, and P. Treutlein, Quantum metrology with nonclassical states of atomic ensembles, Rev. Mod. Phys. \textbf{90}, 035005 (2018).

\bibitem{MCRechtsmanOptica2016}
M. C. Rechtsman, Y. Lumer, Y. Plotnik, A. Perez-Leija, A. Szameit, and M. Segev, Topological protection of photonic path entanglement, Optica \textbf{3}, 925 (2016).
\bibitem{MWangNanophotonics2019}
M. Wang, C. Doyle, B. Bell, M. J. Collins, E. Magi, B. J. Eggleton, M. Segev, and A. Blanco-Redondo, Topologically protected entangled photonic states, Nanophotonics \textbf{8}, 1327 (2019).
\bibitem{KMonkmanPRR2020}
K. Monkman and J. Sirker, Operational entanglement of symmetry-protected topological edge states, Phys. Rev. Res. \textbf{2}, 043191 (2020).
\bibitem{JXHanPRA2021}
J.-X. Han, J.-L. Wu, Y. Wang, Y. Xia, Y.-Y. Jiang, and J. Song, Large-scale Greenberger-Horne-Zeilinger states through a topologically protected zero-energy mode in a superconducting qutrit-resonator chain, Phys. Rev. A \textbf{103}, 032402 (2021).

\bibitem{JLTambascoSA2018}
J. L. Tambasco, G. Corrielli, R. J. Chapman, A. Crespi, O. Zilberberg, R. Osellame, and A. Peruzzo, Quantum interference of topological states of light, Sci. Adv. \textbf{4}, eaat3187 (2018).
\bibitem{CKHongPRL1987}
C. K. Hong, Z. Y. Ou, and L. Mandel, Measurement of subpicosecond time intervals between two photons by interference, Phys. Rev. Lett. \textbf{59}, 2044 (1987).
\bibitem{PGHarper1955}
P. G. Harper, Single band motion of conduction electrons in a uniform magnetic field, Proc. Phys. Soc., London, Sect. A \textbf{68}, 874 (1955).

\bibitem{SHuPRA2020}
S. Hu, Y. Ke, and C. Lee, Topological quantum transport and spatial entanglement distribution via a disordered bulk channel, Phys. Rev. A \textbf{101}, 052323 (2020).

\bibitem{WPSuPRL1979}
W. P. Su, J. R. Schrieffer, and A. J. Heeger, Solitons in Polyacetylene, Phys. Rev. Lett. \textbf{42}, 1698 (1979).
\bibitem{JKAsboth2016}
J. K. Asb\'{o}th, L. Oroszlány, and A. Pályi, A short course on topological insulators, Lect. Notes Phys. \textbf{919}, 1-22 (2016).
  
\bibitem{PhysRevB.86.115112}
C. Fang, M. J. Gilbert, and B. A. Bernevig, Bulk topological invariants in noninteracting point group symmetric insulators, Phys. Rev. B \textbf{86}, 115112 (2012).
\bibitem{PhysRevB.87.035119}
C. Fang, M. J. Gilbert, and B. A. Bernevig, Entanglement spectrum classification of $C_n$-invariant noninteracting topological insulators in two dimensions, Phys. Rev. B \textbf{87}, 035119 (2013).
\bibitem{LFuPRL2012}
L. Fu and C. L. Kane, Topology, delocalization via average
symmetry and the symplectic anderson transition,
Phys. Rev. Lett. \textbf{109}, 246605 (2012).
\bibitem{PhysRevB.89.155424}
I. C. Fulga, B. van Heck, J. M. Edge, and A. R. Akhmerov, Statistical topological insulators, Phys. Rev. B \textbf{89}, 155424 (2014).
\bibitem{PhysRevResearch.2.012067}
A. Agarwala, V. Juri\v{c}i\'{c}, and B. Roy, Higher-order topological insulators in amorphous solids, Phys. Rev. Res. \textbf{2},
012067 (2020).
\bibitem{PhysRevLett.126.206404}
J.-H. Wang, Y.-B. Yang, N. Dai, and Y. Xu, Structuraldisorder-induced second-order topological insulators in three dimensions, Phys. Rev. Lett. \textbf{126}, 206404 (2021).
\bibitem{PhysRevX.13.031016}
R. Ma and C. Wang, Average symmetry-protected topological phases, Phys. Rev. X \textbf{13}, 031016 (2023).
\bibitem{Tao2023}
Y.-L. Tao, J.-H. Wang, and Y. Xu, Average symmetry protected higher-order topological amorphous insulators,
SciPost Phys. \textbf{15}, 193 (2023).

\bibitem{PhysRevB.89.155114}
A. Alexandradinata, X. Dai, and B. A. Bernevig, Wilsonloop characterization of inversion-symmetric topological insulators, Phys. Rev. B \textbf{89}, 155114 (2014).
\bibitem{PhysRevA.102.013301}
Z. Lei, Y. Deng, and C. Lee, Symmetry-protected topological phase for spin-tensor-momentum-coupled ultracold atoms, Phys. Rev. A \textbf{102}, 013301 (2020).
\bibitem{PhysRevB.103.024205}
S. Velury, B. Bradlyn, and T. L. Hughes, Topological crystalline phases in a disordered inversion-symmetric
chain, Phys. Rev. B \textbf{103}, 024205 (2021).

\bibitem{YKePRA2017}
Y. Ke, X. Qin, Y. S. Kivshar, and C. Lee, Multiparticle Wannier states and Thouless pumping of interacting bosons, Phys. Rev. A \textbf{95}, 063630 (2017).
%
WalterNP2023
\bibitem{WLiuPRR2023}
W. Liu, S. Hu, L. Zhang, Y. Ke, and C. Lee, Correlated topological pumping of interacting bosons assisted by Bloch oscillations, Phys. Rev. Research \textbf{5}, 013020, (2023).
%
\bibitem{WalterNP2023}
A.-S. Walter, Z. Zhu, M. Gächter, J. Minguzzi, S. Roschinski, K. Sandholzer, K. Viebahn, and T. Esslinger, Quantization and its breakdown in a Hubbard–Thouless pump, Nat. Phys. \textbf{19}, 1471, (2023).
%


\bibitem{ZilberbergNature2018}
O. Zilberberg, S. Huang, J. Guglielmon, M. Wang, K. P. Chen, Y. E. Kraus, and Mikael C. Rechtsman, Photonic topological boundary pumping as a probe of 4D quantum Hall physics, Nature \textbf{533}, 59 (2018).


\end{thebibliography}
\end{document}